\documentclass[reprint,amsmath,amssymb,aps,superscriptaddress]{revtex4-1}

\usepackage{graphicx}
\usepackage{dcolumn}
\usepackage{soul}
\usepackage[normalem]{ulem}
\usepackage{bm}
\usepackage[caption=false]{subfig}

\newcommand{\ket}[1]{\arrowvert #1 \rangle}

\begin{document}

\preprint{APS/123-QED}

\title{Coherent Manipulation of Orbital Feshbach Molecules of Two-Electron Atoms}

\affiliation{Dipartimento di Fisica e Astronomia, Universit\`a degli Studi di Firenze, I-50019 Sesto Fiorentino, Italy}
\affiliation{LENS European Laboratory for Nonlinear Spectroscopy, I-50019 Sesto Fiorentino, Italy}
\affiliation{INO-CNR Istituto Nazionale di Ottica del CNR, Sezione di Sesto Fiorentino, I-50019 Sesto Fiorentino, Italy}
\affiliation{INFN Istituto Nazionale di Fisica Nucleare, Sezione di Firenze, I-50019 Sesto Fiorentino, Italy}

\author{G.~Cappellini}
\thanks{cappellini@lens.unifi.it}
\affiliation{Dipartimento di Fisica e Astronomia, Universit\`a degli Studi di Firenze, I-50019 Sesto Fiorentino, Italy}
\affiliation{INO-CNR Istituto Nazionale di Ottica del CNR, Sezione di Sesto Fiorentino, I-50019 Sesto Fiorentino, Italy}
\affiliation{INFN Istituto Nazionale di Fisica Nucleare, Sezione di Firenze, I-50019 Sesto Fiorentino, Italy}

\author{L.~F.~Livi}
\affiliation{Dipartimento di Fisica e Astronomia, Universit\`a degli Studi di Firenze, I-50019 Sesto Fiorentino, Italy}
\affiliation{INO-CNR Istituto Nazionale di Ottica del CNR, Sezione di Sesto Fiorentino, I-50019 Sesto Fiorentino, Italy}
\affiliation{INFN Istituto Nazionale di Fisica Nucleare, Sezione di Firenze, I-50019 Sesto Fiorentino, Italy}

\author{L.~Franchi}
\affiliation{Dipartimento di Fisica e Astronomia, Universit\`a degli Studi di Firenze, I-50019 Sesto Fiorentino, Italy}
\affiliation{INFN Istituto Nazionale di Fisica Nucleare, Sezione di Firenze, I-50019 Sesto Fiorentino, Italy}

\author{D.~Tusi}
\affiliation{Dipartimento di Fisica e Astronomia, Universit\`a degli Studi di Firenze, I-50019 Sesto Fiorentino, Italy}

\author{D.~Benedicto Orenes}
\affiliation{INO-CNR Istituto Nazionale di Ottica del CNR, Sezione di Sesto Fiorentino, I-50019 Sesto Fiorentino, Italy}

\author{M.~Inguscio}
\affiliation{Dipartimento di Fisica e Astronomia, Universit\`a degli Studi di Firenze, I-50019 Sesto Fiorentino, Italy}
\affiliation{LENS European Laboratory for Nonlinear Spectroscopy, I-50019 Sesto Fiorentino, Italy}
\affiliation{INO-CNR Istituto Nazionale di Ottica del CNR, Sezione di Sesto Fiorentino, I-50019 Sesto Fiorentino, Italy}

\author{J.~Catani}
\affiliation{INO-CNR Istituto Nazionale di Ottica del CNR, Sezione di Sesto Fiorentino, I-50019 Sesto Fiorentino, Italy}
\affiliation{LENS European Laboratory for Nonlinear Spectroscopy, I-50019 Sesto Fiorentino, Italy}
\affiliation{INFN Istituto Nazionale di Fisica Nucleare, Sezione di Firenze, I-50019 Sesto Fiorentino, Italy}

\author{L.~Fallani}
\affiliation{Dipartimento di Fisica e Astronomia, Universit\`a degli Studi di Firenze, I-50019 Sesto Fiorentino, Italy}
\affiliation{LENS European Laboratory for Nonlinear Spectroscopy, I-50019 Sesto Fiorentino, Italy}
\affiliation{INO-CNR Istituto Nazionale di Ottica del CNR, Sezione di Sesto Fiorentino, I-50019 Sesto Fiorentino, Italy}
\affiliation{INFN Istituto Nazionale di Fisica Nucleare, Sezione di Firenze, I-50019 Sesto Fiorentino, Italy}

\begin{abstract}

Ultracold molecules have experienced increasing attention in recent years. Compared to ultracold atoms, they possess several unique properties that make them perfect candidates for the implementation of new quantum-technological applications in several fields, from quantum simulation to quantum sensing and metrology. In particular, ultracold molecules of two-electron atoms (such as strontium or ytterbium) also inherit the peculiar properties of these atomic species, above all the possibility to access metastable electronic states via direct excitation on optical clock transitions with ultimate sensitivity and accuracy. \par
In this paper we report on the production and coherent manipulation of molecular bound states of two fermionic $^{173}$Yb atoms in different electronic (orbital) states $^1$S$_0$ and $^3$P$_0$ in proximity of a scattering resonance involving atoms in different spin and electronic states, called \textit{orbital Feshbach resonance}. We demonstrate that orbital molecules can be coherently photoassociated starting from a gas of ground-state atoms in a three-dimensional optical lattices by observing several photoassociation and photodissociation cycles. We also show the possibility to coherently control the molecular internal state by using Raman-assisted transfer to swap the nuclear spin of one of the atoms forming the molecule, thus demonstrating a powerful manipulation and detection tool of these molecular bound states. Finally, by exploiting this peculiar detection technique we provide first information on the lifetime of the molecular states in a many-body setting, paving the way towards future investigations of strongly interacting Fermi gases in a still unexplored regime.
\end{abstract}

\maketitle

\section{Introduction}

The experimental study of ultracold molecules gained significant momentum in the last decade, as molecules can considerably extend the range of application of ultracold quantum technology to a wealth of different phenomena and regimes hardly accessible to atoms \cite{carr2009}. For instance, simple molecules are a valuable resource for high-precision spectroscopy, as they can exhibit much increased sensitivities (with respect to atoms) to the high-precision measurements of fundamental constants (e.g. the electron-to-proton mass ratio) or to the search of new physics and elusive quantum effects \cite{chin2009,safronova2018}. From a many-body perspective, interactions between molecules show a significantly richer physics than that of neutral atoms and great effort has been recently devoted to the realization of ultracold gases of polar molecules \cite{moses2015,reich2017,park2015,molony2014,guo2016,rvachov2017,seesselberg2018}, eventually reaching the Fermi-degenerate regime \cite{demarco2018}. The interest for such systems relies in the possibility of exploiting the strong dipole-dipole interaction connected with the permanent electric dipole of molecules, that could allow the observation of new strongly correlated quantum phases \cite{micheli2006,yan2013}. Furthermore, the study of interactions between molecules and/or atoms prepared in well controlled quantum states enabled first experimental studies in the emerging field of ultracold quantum chemistry \cite{ospelkaus2010,mcdonald2016}. 

While extraordinary progress has been made in very recent years in the cooling of room-temperature stable molecules \cite{zeppenfeld2012,barry2014,truppe2017,hemmerling2016} the controlled synthesis of molecular bound states starting from ensembles of ultracold atoms is still a leading direction of research. Most of the experimental activity in this context focused on alkali atoms, where magnetic Feshbach resonances can be used to adjust the interaction between cold atoms and associate them in weakly-bound dimers with tunable binding energy \cite{chin2010}.

Extending this study to molecules made by alkaline-earth atoms (or alkaline-earth-like atoms such as ytterbium) would open completely new scenarios. Indeed, the two-electron structure of these atoms entails the existence of low-lying (spin-triplet) metastable electronic states, that can be excited from the (spin-singlet) ground state on narrow intercombination optical transitions. Among these, the doubly-forbidden transition connecting the ground state $^1$S$_0$ and the excited state $^3$P$_0$ (with sub-Hz natural linewidth) is well known in the context of frequency metrology, as it is used for realizing the most precise and accurate atomic clocks to date at the $10^{-18}$ level \cite{bloom2014,ushijima2015,mcgrew2018}. The coherent manipulation of molecular states made by two-electron atoms could disclose new opportunities in this context, with the possible development of sub-Hz-linewidth molecular atomic clocks with enhanced ``sensing" capabilities \cite{Borkowski2018}. Previous experimental works in a related context include the production and manipulation of cold Sr molecules on the $^1$S$_0$--$^3$P$_1$ transition  \cite{mcguyer2015,mcdonald2016,mcguyer2015b} and early studies on the photoassociation of molecular bound states of Yb atoms in metastable states \cite{Hofer2015,Taie2016,Takasu2017}. However, no coherent manipulation of molecules made by two-electron atoms in “clock” states $^1$S$_0$--$^3$P$_0$ of metrological relevance has been reported yet.

\begin{figure}[t!] 
\includegraphics[width=\columnwidth]{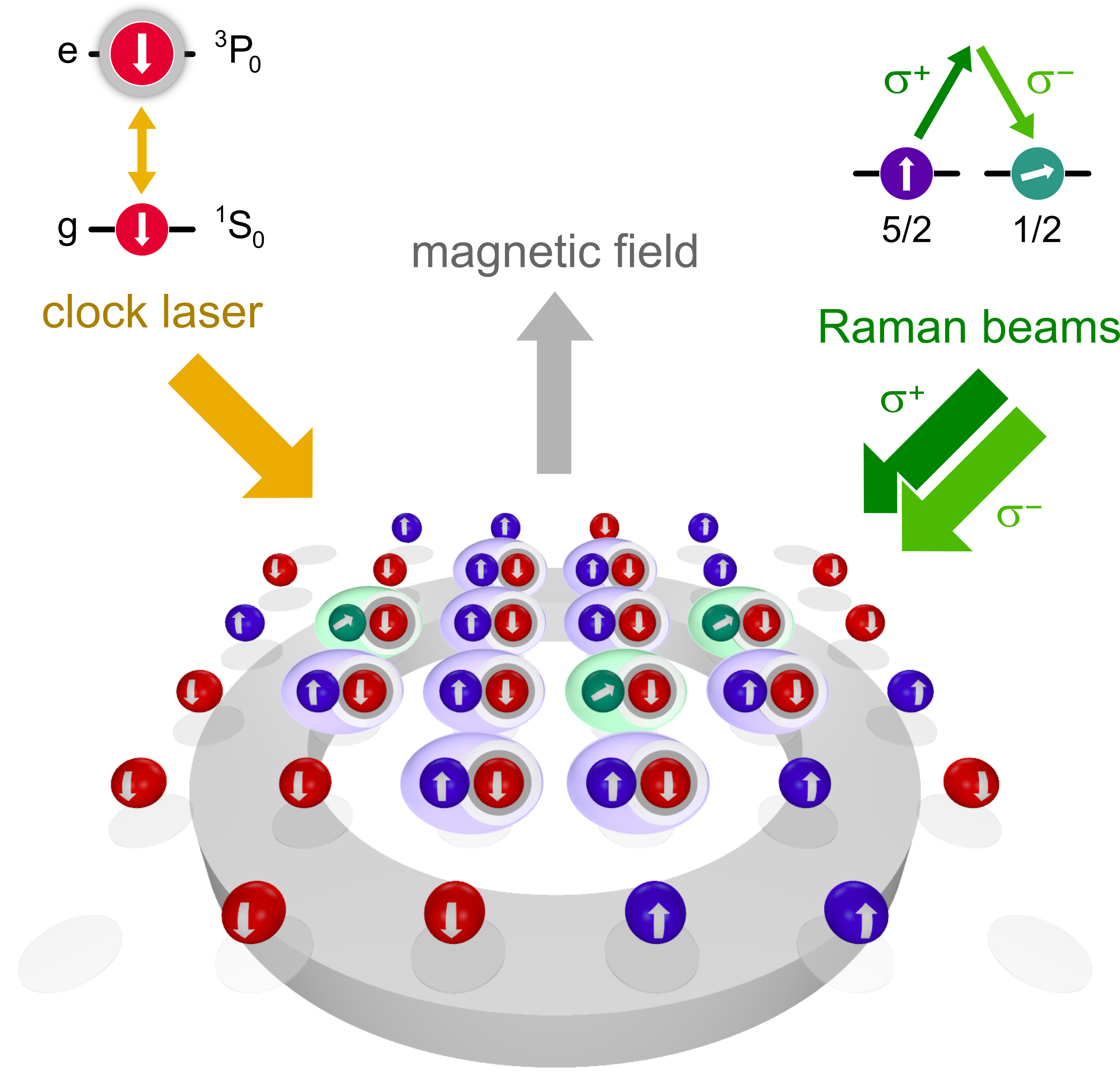}
\caption{Sketch of the experimental system. Ultracold $^{173}$Yb atoms are trapped in optical lattices (of variable dimensionality depending on the specific experiments). The electronic (orbital) state is manipulated by an ultranarrow clock laser, while a pair of Raman laser beams controls the nuclear spin state. Interactions between atoms in different internal states are controlled with a magnetic field tuned in proximity of the Orbital Feshbach Resonance.}  \label{fig:intro_sketch}
\end{figure}

The manipulation of those ``clock" molecules is particularly favoured in $^{173}$Yb, as the molecular levels can be controlled thanks to a peculiar kind of Feshbach resonance, called Orbital Feshbach Resonance (OFR) \cite{Zhang2015}. This resonance, based on the interorbital spin-exchange interaction mechanism arising in two-electron fermions with a ``clock" (orbital) degree of freedom \cite{Cappellini2014,Scazza2014}, was observed for the first time in Refs. \cite{Pagano2015,Hofer2015}, thanks to the favourable scattering properties of $^{173}$Yb. The interest for the OFR of $^{173}$Yb is manifold. First, it is characterized by unusually small energy scales, both for the energy of the resonant bound states and for the separation between the coupled scattering channels \cite{cheng2017}, lying both in the same kHz-range as the Fermi energy. This property results in intriguing many-body physics. For instance, it was suggested that the resulting many-body superfluid state has to be described in terms of two coupled order parameters \cite{Zhang2015}, opening the door to study the physics of two-gap superconductors, with the prediction of new collective excitations and the emergence of the long-sought massive Leggett mode \cite{He2016,Zhang2017}.
Furthermore, the OFR has the unique character of a narrow Feshbach resonance featuring a broad magnetic field tunability, which could allow the study of the BEC-BCS crossover in unexplored regimes \cite{xu2016}. In addition, as it was demonstrated in previous experimental works \cite{Livi2016,Kolkowitz2017,Japan}, it is possible to use ultranarrow optical transitions to create single-photon spin-orbit coupling in a pure two-level system, thus avoiding the detrimental heating observed in multi-level configurations in alkali systems. The realization of robust spin-orbit coupling in strongly-interacting Fermi gases with tunable interactions could lead to yet unobserved effects, such as the emergence of ultracold topological phases with Majorana zero-energy excitation modes \cite{liu2012a,lui2012b,zhai2015,iemini2017}. However, working with this peculiar Feshbach resonance introduces significant experimental challenges that are connected with the very small energy scales responsible for the rich physics discussed above.

In this manuscript, we report on the coherent production of weakly-bound molecules made by $^1$S$_0$--$^3$P$_0$ $^{173}$Yb atom pairs in proximity of the OFR, and on the coherent optical manipulation of their internal state. Figure 1 shows a sketch of the system under investigation and of the main experimental tools that we use for the control of the molecular state. The manuscript is organized as follows. In Section II we recall the main properties of the OFR in two-electron fermions. In Section III we demonstrate the possibility of coherent photoassociation of molecules via the optical clock transition, showing high-fidelity photoassociation and photodissociation cycles. In section IV we show that, by exploiting the SU(N) character of the interactions in two-electron atoms, we can use coherent Raman coupling to control the nuclear spin composition of the molecules, as a powerful manipulation and detection tool. In section V we demonstrate the long lifetime of isolated molecules and provide first measurements on the lifetime of the molecular states in a many-body setting. Section VI will be devoted to conclusions.

\section{Orbital Feshbach Resonance in $^{173}$Yb} \label{sec:ofr}

In this Section we briefly review the properties of the Orbital Feshbach Resonance exhibited by $^{173}$Yb atoms (see Refs. \cite{Zhang2015,Pagano2015,Hofer2015} for more details). This scattering resonance arises from the existence of two different degrees of freedom, an electronic (orbital) state and a nuclear spin state, which are coupled by an interorbital spin-exchange interaction \cite{Cappellini2014,Scazza2014}. We consider two atoms in two different orbital states $|g\rangle$ $=$ $^1$S$_0$ and $|e\rangle$ $=$ $^3$P$_0$, and two different nuclear-spin projection states $|$$\uparrow$$\rangle$ and $|$$\downarrow$$\rangle$ chosen arbitrarily out of the $I=5/2$ nuclear spin manifold. Atom-atom interactions are described by two different short-range scattering potentials corresponding to the two exchange-symmetrized two-body states $|eg^\pm\rangle = |g$$\uparrow, e$$\downarrow \rangle \mp |g$$\downarrow, e$$\uparrow \rangle$. A nonzero magnetic field introduces a coupling between the $|eg^\pm\rangle$ states and defines the long-distance asymptotic scattering channels $|c\rangle = |g\downarrow, e\uparrow \rangle$ (closed channel) and $|o\rangle = |g\uparrow, e\downarrow \rangle$ (open channel). As in ordinary Feshbach resonances, the two-body $s-$wave scattering length diverges when the scattering state for two atoms colliding in the $|o\rangle$ channel becomes quasi-resonant with a bound state of the $|c\rangle$ potential. In the specific case of $^{173}$Yb the resonance occurs with the least bound state of $|c\rangle$, that has a binding energy of only a few kHz. This very small value allows the possibility of accessing the resonance regime even if the magnetic field tuning is strongly suppressed by the purely nuclear character of the atomic spin (with a differential magnetic sensitivity of the two channels of $\approx 110$ Hz/G $\times$ $\Delta m$, where $\Delta m$ is the difference in the spin projection quantum numbers of the $|$$\uparrow$$\rangle$ and $|$$\downarrow$$\rangle$ states).

\begin{figure}[t!] 
\includegraphics[width=\columnwidth]{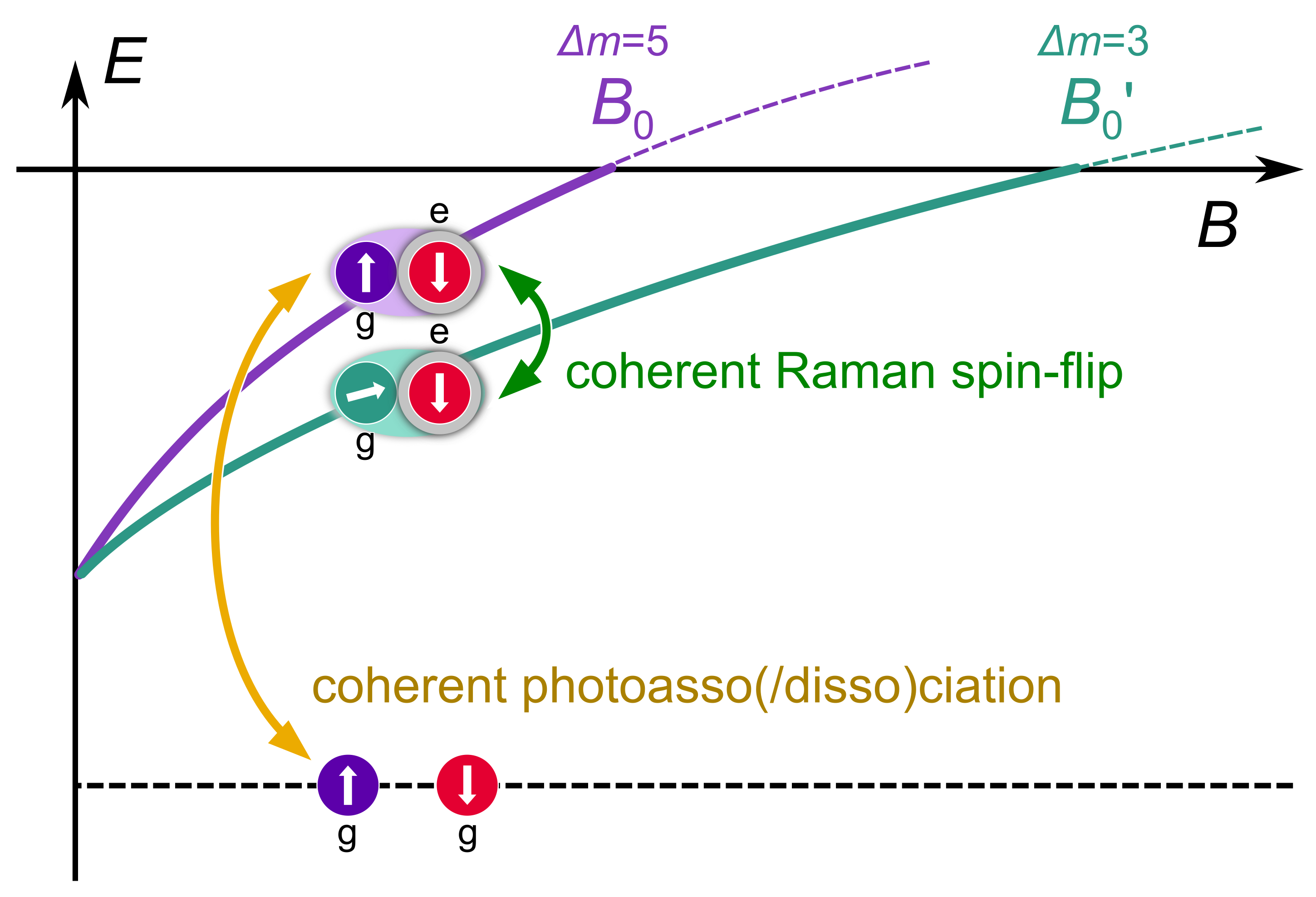}
\caption{Sketch of the processes considered in this work and corresponding energy levels. The solid lines show the binding energies for $^{173}$Yb $|e\rangle$--$|g\rangle$ atom pairs with different values of $\Delta m$ (difference between the nuclear spin projection quantum numbers) and $B_0$ ($B_0'$) denotes the position of the OFR center. The arrows show the different laser transitions used in the experiment for molecular production/dissociation and control of the internal state.} \label{fig:intro_scheme}
\end{figure}

It is indeed because of the last property discussed above that the OFR exhibits the character of a narrow Feshbach resonance \cite{xu2016}, while keeping a very broad tunability in terms of magnetic field accessibility. As an example, for $^{173}$Yb atoms in the $|$$\uparrow\rangle=|+5/2\rangle$, $|$$\downarrow\rangle=|-5/2\rangle$ states, the resonance is located at $B_0 \simeq 40$ G, with a zero-crossing of the scattering length at around $B \simeq 400$ G. As mentioned before, this unusual feature could open to the first experimental investigation of the BEC-BCS crossover close to a narrow Feshbach resonance, allowing to overcome the strict magnetic-field stability requirements for ordinary Feshbach resonances in alkali gases.

A natural direction for the experimental investigation of these phenomena is to follow the path of fermionic alkali gases, where the first experiments aimed at the formation of ultracold fermionic molecules on the BEC side of the resonance, followed by measurements of their stability \cite{cubizolles2003} and by their Bose-condensation in a superfluid state \cite{greiner2003,jochim2003,zwierlein2003}.

However, the same reason that makes the system interesting (the small energy scales), also precludes the possibility of a direct application of the experimental tools that were developed to control and detect Feshbach molecules of alkali atoms. For instance, already a direct imaging of the molecules is prohibited by their binding energy being much smaller than the linewidth of the imaging transition. Therefore different tools need to be implemented (we will discuss them in Section IV).

We conclude this section by mentioning that the OFR follows a very simple scaling relation with the magnetic field $B$ and the difference in nuclear spin projection quantum numbers $\Delta m$ of the interacting atoms. As a matter of fact, the scattering length $a$ is a universal function of $B \Delta m$: $a(B,\Delta m)=\tilde a (B \Delta m)$. This property stems from the SU(N) character of the interactions in two-electron fermions \cite{gorshkov2010,Pagano2015,Hofer2015}, namely the independence of the scattering potentials on the nuclear spin state, which cause $\Delta m$ to affect only the Zeeman splitting between the two scattering channels (and not their individual properties). This behavior is exemplified by the binding energies on the BEC side of the resonance, that are sketched in Fig. 2 for two different atom pairs with $\Delta m=5$ and $\Delta m =3$: while the two binding energies at zero field are degenerate, they differ at a finite magnetic field according to the same scaling $E_b(B,\Delta m)=\tilde E_b (B \Delta m)$. We will take advantage of this scaling in the Raman-based spectroscopy discussed in Section IV.

\section{Coherent association of orbital molecules} \label{sec:photoass}

In this section we show how molecular bound states of $^{173}$Yb atoms in different electronic states ($\ket{g}$ and $\ket{e}$) can be produced by means of coherent photoassociation pulses of clock laser light in a 3D optical lattice. In this configuration, the molecules are localized in individual lattice sites, preventing short-term inelastic losses and allowing for an easier study of their basic properties, as well as for the observation of long-lived coherent photoassociation and photodissociation cycles of orbital molecules. Moreover, the presence of the lattice confinement increases the stability of the dimer \cite{Stoferle2006,Thalhammer2006,Ospelkaus2006}, allowing for the existence of bound states at magnetic field values higher than the center of the free-space Feshbach resonance, located at approximately 40 G \cite{Pagano2015,Hofer2015}, as well as for an increase of the low-field binding energy. \par 
The experiment is initiated by preparing a Fermi gas of approximately 70$\times10^3$ ytterbium atoms at a temperature $T = 0.25\ T_F$ (with $T_F$ the Fermi temperature) in the $m = \pm 5/2$ nuclear spin states. The gas is obtained by evaporating in a 3D optical trap at 1064 nm with final trapping frequency of $2\pi\times$(93,73,86) Hz. A 3D optical lattice operating at the magic wavelength $\lambda_L = 2\pi /k_L=$ 759 nm is adiabatically ramped up to a depth of $V_0=s E_r$, with $s = 15$ and where $E_r = \hbar^2 k_L^2/2M$ is the recoil energy of the lattice, with $M$ being the atomic mass. The dipole trap at 1064 nm is then turned off with a 1-s-long linear ramp and the lattices are subsequently increased in 1 ms to a final lattice depth $s_f$ between 15 and 30 (sufficient to freeze the translational degree of freedom) depending on the specific experiment. This lattice loading procedure is adjusted to maximize the number of lattice sites occupied by a pair of $m=\pm5/2$ atoms. \par
\begin{figure} 
\includegraphics[width=\columnwidth]{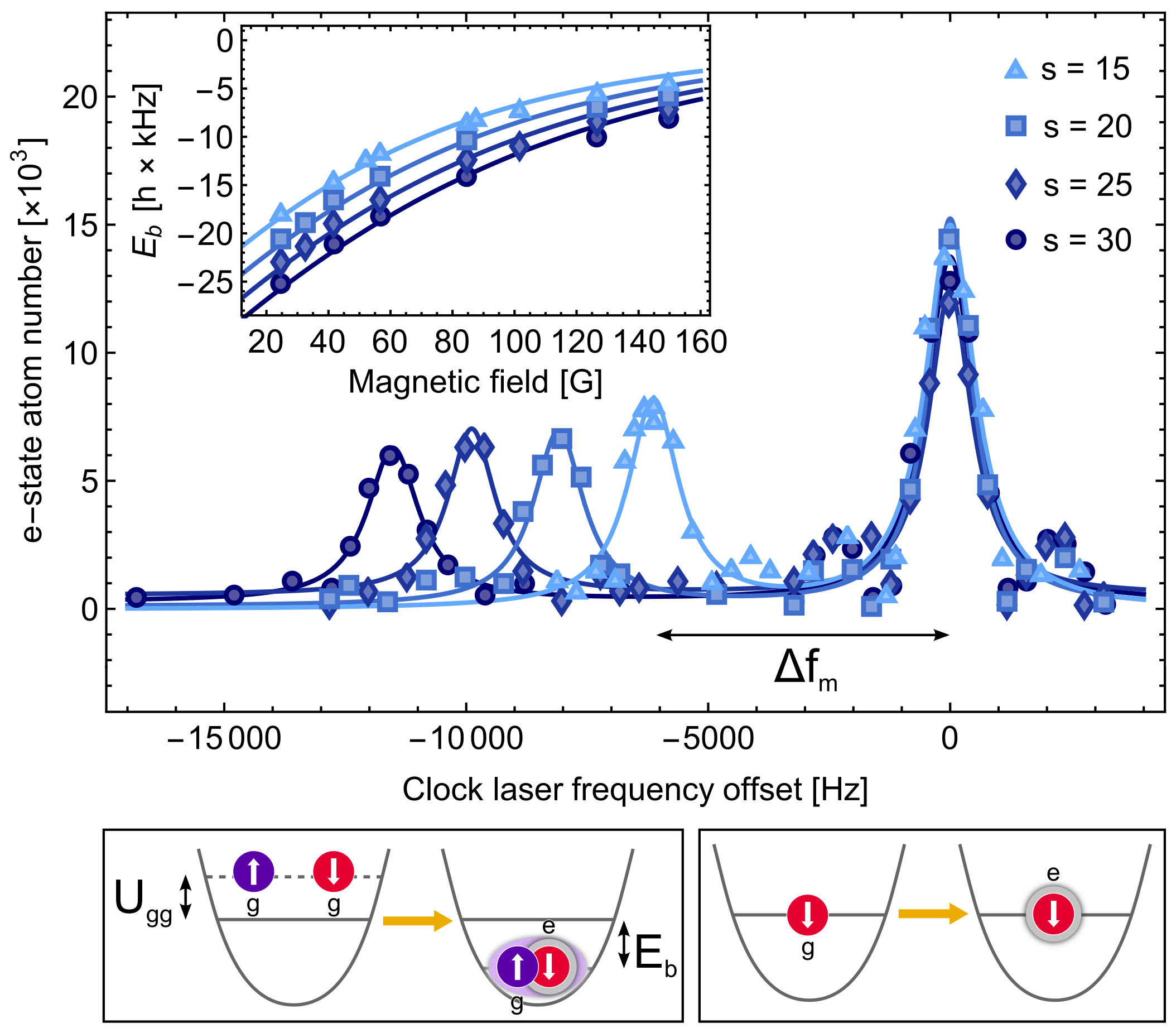}
\caption{Clock laser spectroscopy of a $m=\pm5/2$ sample of $^{173}$Yb in a 3D lattice for different values of lattice depth and 150 G. The zero-frequency reference corresponds to the excitation of individual $m=-5/2$ atoms in singly-occupied lattice sites, while the peaks at lower frequency correspond to the photoassociation process of $\ket{g}-\ket{e}$ bound states. The inset reports the value of the binding energy as a function of the magnetic field for different lattice depths and the solid lines are a fit with the theoretical model (see text for details).} \label{fig:photoass_spec}
\end{figure}

In order to produce orbital molecules we excite the $\ket{g}$ $\rightarrow$ $\ket{e}$ clock transition at a specific magnetic field $B$ by probing the atomic sample in the Lamb-Dicke regime with pulses of $\pi$-polarized clock-laser light generated by a frequency-doubled solid-state laser locked to an ultrastable optical cavity with a $<50$ Hz linewidth \cite{Cappellini2015}, similarly to the experiment performed in Ref. \cite{Hofer2015}. In doubly occupied sites, different two-particle states can be excited at different clock laser frequencies. While two-particle states of fermions with repulsive interactions have already been investigated both in Yb \cite{Cappellini2014,Scazza2014} and Sr \cite{Campbell2017} atoms, here we perform a detailed study of the molecular branch located in a lower frequency region of the spectrum, that was earlier observed in Ref. \cite{Hofer2015}. Fig. \ref{fig:photoass_spec} reports a series of typical clock spectroscopy spectra at $B = 150$ G for different depths of the 3D lattice obtained by measuring the number of atoms in the excited $\ket{e}$ state after the clock laser pulse. This is done by removing the residual ground-state atoms after the clock excitation with a resonant imaging pulse, after releasing the cloud from the lattice and pumping with a 1389 nm laser the $\ket{e}$ atoms into the $^3$D$_1$ state. From there they then decay, through the intermediate $^3$P$_1$ level, to the ground state where they can be detected with the usual imaging procedure \cite{Franchi2017}. In the spectra of Fig. \ref{fig:photoass_spec}, the zero-frequency reference of the horizontal axis is the single particle $\ket{g}$ $\rightarrow$ $\ket{e}$ transition for individual atoms in the $m = -5/2$ state. This transition frequency only depends on the magnetic field and not on the lattice depth (residual light shifts induced by the lattice light are negligible at the resolution level of this experiment) and corresponds to the total energy of a pair of non-interacting $\ket{e,-5/2}$ and $\ket{g,5/2}$ atoms, i.e. the threshold energy of the orbital Feshbach resonance. In every dataset, the peak located at lower frequency corresponds to the excitation of a $\ket{g}-\ket{g}$ pair to a $\ket{g,+5/2;e,-5/2}$ orbital molecular state characterized by a binding energy $E_b$. In the following the notation $\ket{g;e}$ could also be used for simplicity to indicate a generic molecular state of two atoms in $\ket{g}$ and $\ket{e}$ states. By measuring the distance $\Delta f_{m}$ between the single particle peak and the molecular photoassociation peak, the binding energy can be determined as $E_b = h\Delta f_{m} - U_{gg}$, where $h$ is the Planck constant and $U_{gg}$ is the repulsive Hubbard interaction energy of the two initial ground-state $m=\pm 5/2$ atoms calculated using the scattering length values reported in Ref. \cite{Kitagawa2008}. 

With this approach, we measured the binding energy $E_b$ of a $\ket{g,+5/2;e,-5/2}$ molecule as a function of the magnetic field $B$ for several values of the lattice depth ranging from $s_f = 15$ to $s_f = 30$. The results of these measurements are reported in the inset of Fig. \ref{fig:photoass_spec}. As anticipated before, it should be noted that a molecular state exists even at magnetic field values higher than the position of the free-space resonance around 40 G due to lattice confinement. The solid lines in the inset are a global fit of the full experimental dataset with the theoretical model discussed in Ref. \cite{Hofer2015} that derives the molecular binding energy by extending the problem of two atoms interacting in a harmonic trap with a single interaction channel  \cite{Busch1998} to the case of a singlet and triplet interaction channels coupled by a magnetic field. Anharmonic corrections due to the finite lattice depth have been evaluated numerically by diagonalizing the full Hamiltonian of the problem considering the coupling between relative and center-of-mass motion up to the fourth lattice band (see supplementary materials of Ref. \cite{Cappellini2014}). From the fit, in which only the spin triplet scattering length $a_{eg}^+$ is left as a free parameter, we estimate $a_{eg}^+=1894(18)\,a_0$ where $a_0$ is the Bohr radius, a value in agreement with the one determined in Ref. \cite{Hofer2015}.\par
\begin{figure} 
\includegraphics[width=\columnwidth]{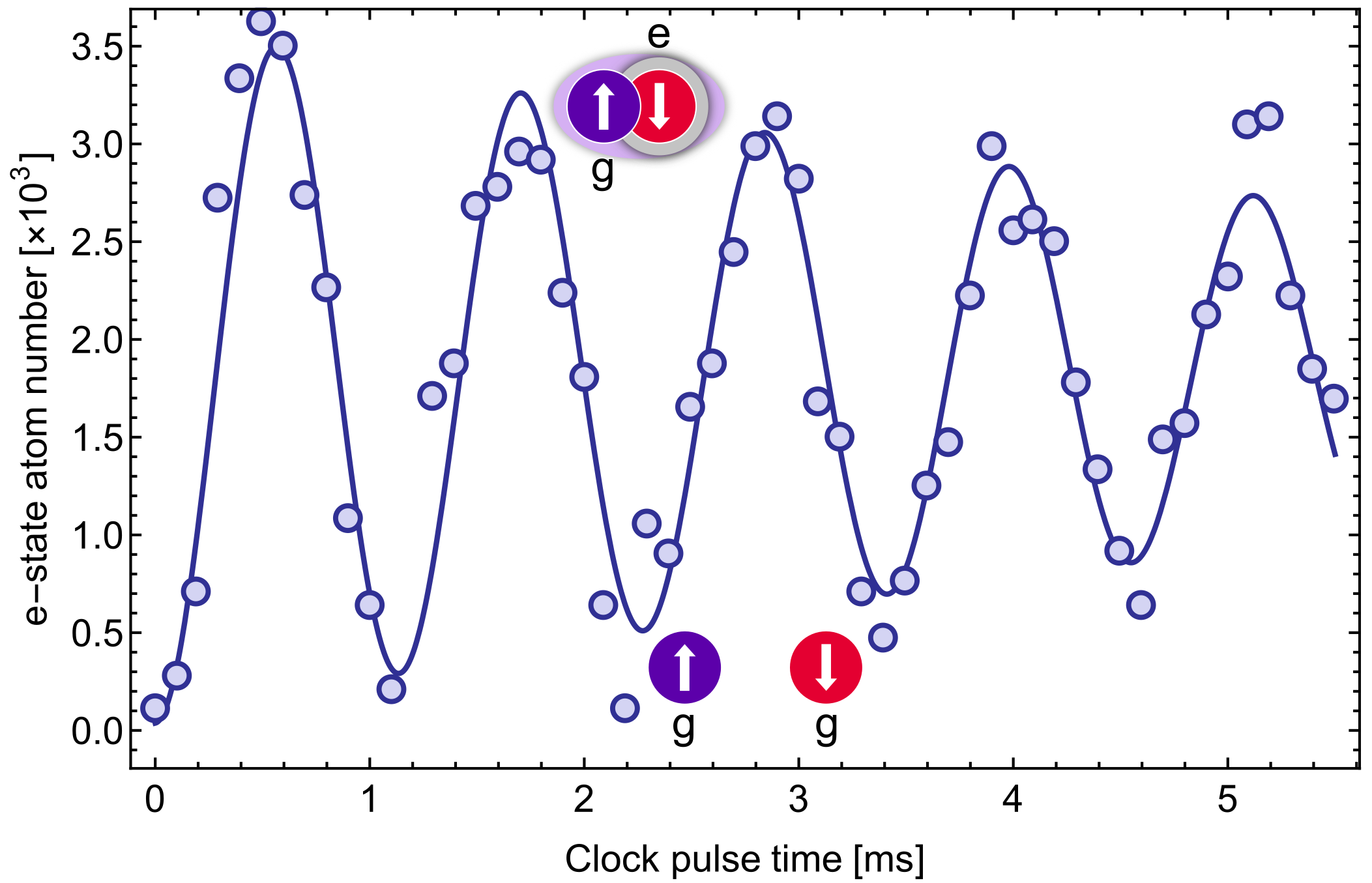}
\caption{Rabi oscillation on a $\ket{g}-\ket{e}$ photoassociation peak performed at a magnetic field $B=150$ G and 3D lattice depth $s_f=30$. The points are the experimental data and the solid curve is a fit with a damped sinusoidal function. Several periods of photoassociation and photodissociation cycles can be observed. No error bars are present since every point is the result of a single acquisition.} \label{fig:photoass_rabi}
\end{figure}

In a subsequent experiment we demonstrate the capability to drive a coherent molecule photoassociation process \cite{Taie2016,Fu2014}. In this experiment, the clock laser frequency is kept constant at the photoassociation peak value and the sample of atoms confined in the 3D lattice is probed with clock laser pulses of increasing duration. Fig. \ref{fig:photoass_rabi} reports a typical oscillation at a field $B=150$ G and a 3D lattice depth $s_f=30$ between pairs of free $\ket{g}-\ket{g}$ atoms and orbital $\ket{g;e}$ bound states in doubly-occupied lattice sites, obtained by detecting the number of $\ket{e}$-state atoms with the same technique used to measure the spectra of Fig. \ref{fig:photoass_spec}. The points are the experimental data and the solid line is the result of a fit to the data with a damped sinusoidal function. Several cycles of photoassociation and photodissociation can be observed, and the ratio between the frequency $\Omega$ of this oscillation and its single-particle counterpart $\Omega_0$ driven with the same clock laser intensity is a direct measurement of the Franck-Condon factor, which is $\mathcal{F}=\Omega/\Omega_0 = 0.81(6)$ at this magnetic field value. The possibility to coherently excite pairs of atoms to the bound state is a remarkable feature, that in our experiment is fundamental to maximize the number of orbital molecules in the sample (a goal that is achieved by shining a $\pi$-pulse of clock laser light at the photoassociation frequency), but is in general interesting for the implementation of new experimental protocols involving the coherent manipulation of two-particle states, e.g. for high-precision spectroscopy \cite{Borkowski2018}.

\section{Coherent spin manipulation of orbital molecules}

In this section we show how orbital $\ket{g;e}$ molecules can be coherently manipulated and detected by exploiting Raman transitions acting on the ground-state nuclear spin degree of freedom of the molecular state. We start by preparing a sample of $^{173}$Yb atoms in the $m=\pm5/2$ states in a 3D lattice as reported in the previous section. Atoms in doubly occupied sites are converted into orbital molecules by performing a clock laser $\pi$-pulse resonant with the photoassociation frequency, so that the sample consists of a 3D array of $\ket{g,+5/2;e,-5/2}$ orbital molecules and individual atoms either in the $m=-5/2$ or $m=+5/2$ state. We then perform Raman spectroscopy on the sample with a pair of beams with frequencies $f$ and $f+\Delta f$ detuned by 1756 MHz from the $\ket{^1\mathrm{S}_0\ (F=5/2)}\rightarrow\ket{^3\mathrm{P}_1\ (F=7/2)}$ intercombination line ($\sim 10^4$ times the transition natural linewidth $\Gamma\simeq180$ kHz), inducing transitions between different nuclear spin states of the ground state manifold. The two Raman beams are co-propagating, so that no momentum kick is imparted to the sample during the Raman process, and their polarization is an equal superposition of $\sigma^+$ and $\sigma^-$, thus inducing two possible two-photon processes, namely $\ket{-5/2}\rightarrow\ket{-1/2}$ and $\ket{+5/2}\rightarrow\ket{+1/2}$ transitions. \par

\begin{figure} 
\includegraphics[width=\columnwidth]{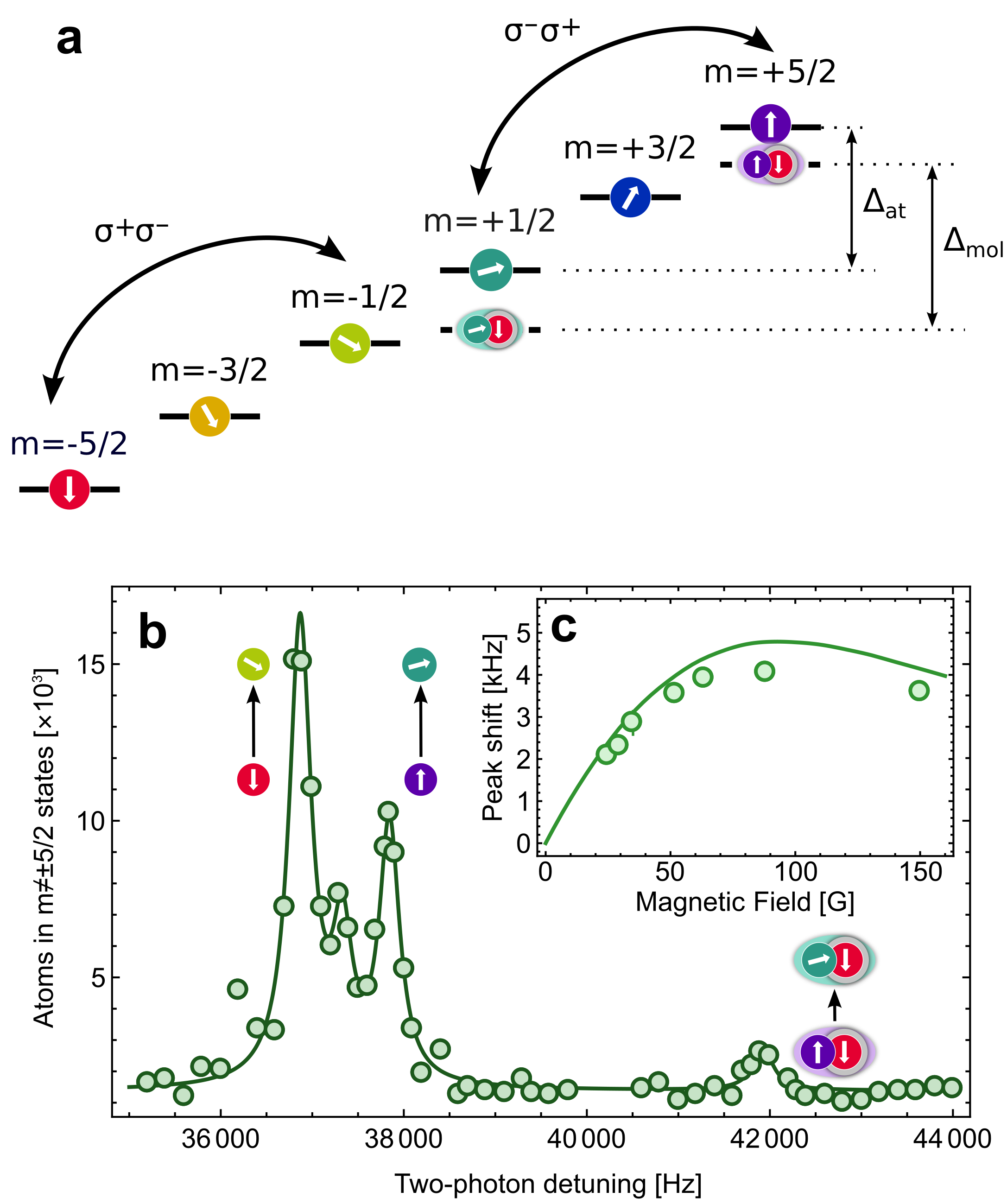}
\caption{\textbf{a.} Scheme of the two-photon Raman transitions connecting different nuclear spin states. The energy $\Delta_{at}$ corresponding to a Raman spin-flip of individual atoms differs from the energy $\Delta_{mol}$ needed to flip the spin of the ground-state atom in the orbital molecule (because of the different  binding energies for different spin states). \textbf{b.} Raman spectroscopy of a sample containing individual atoms with $m=\pm 5/2$ and orbital molecules at $B=88$ G and $s_f = 15$. The peaks at lower frequency correspond to the the spin flipping of individual atoms while the small peak near 42 kHz corresponds to a bound-to-bound transition between orbital molecules made of different spin mixtures. The inset shows the shift of the bound-to-bound transition peak measured as a function of the magnetic field and the solid line is the binding energy difference between the initial and final molecular states calculated from our model (see section \ref{sec:photoass}).} \label{fig:raman_spec}
\end{figure}

Fig. \ref{fig:raman_spec}{\bf b} reports a typical Raman spectrum at a magnetic field of $B=88$ G and a 3D lattice depth of $s_f=15$. In order to perform a zero-background measurement, the stretched states $m=\pm 5/2$ are selectively blasted with pulses resonant with the magnetically sensitive $\ket{^1\mathrm{S}_0}\rightarrow\ket{^3\mathrm{P}_1}$ transition during the time-of-flight. As a result, only atoms having flipped their spin to other nuclear spin states as a consequence of a Raman process are detected. This detection technique is affected by a small background of a few hundreds of atoms in the $m\neq \pm 5/2$ states as a consequence of the imperfect preparation of the initial sample. In the lower range of two-photon detuning $\Delta f$ between 36 and 38 kHz, the spectrum shows two peaks which can be ascribed to the transition of individual $m = \pm 5/2$ atoms in singly occupied sites to the $m = \pm 1/2$ states. Such processes occur when $\Delta f = \Delta _{at}$, where $\Delta_{at} = \Delta m B\times 207$ Hz/G is the splitting induced by the magnetic field $B$ (see Fig. \ref{fig:raman_spec}{\bf a}). While the two transitions should in principle be degenerate, a differential spin-dependent light shift on the nuclear spin states induced by the Raman light \cite{Mancini2015} lifts the degeneracy between them. In particular, the beam frequencies and polarizations employed in the experiment result in the $\ket{-5/2}\rightarrow\ket{-1/2}$ and $\ket{+5/2}\rightarrow\ket{+1/2}$ transitions being located at lower and higher $\Delta f$, respectively. The different height of the two peaks can be ascribed to the fact that part of the $m=-5/2$ particles in the initially balanced sample of ground-state atoms is excited to the $\ket{e}$-state during the molecule creation process. The third smaller peak between the two single-particle transitions can instead be ascribed to resonant four-photon Raman coupling connecting the $m=\pm 5/2$ states to $m=\mp 3/2$ states. \par
The most relevant feature of the spectrum in Fig. \ref{fig:raman_spec} is the small peak located at higher $\Delta f$ around 42 kHz. This resonance is connected to a process involving the orbital molecules of the sample, since it disappears when molecules are not present in the sample. We ascribe this peak to a bound-to-bound transition process in which the Raman beams act on the ground-state component of the molecules flipping its spin and transforming a $\ket{g,+5/2;e,-5/2}$ molecule into a $\ket{g,+1/2;e,-5/2}$ molecule. The position of this peak is shifted with respect to the single-particle peak(s) due to the different binding energy between the initial and final bound state as a consequence of the binding energy scaling described in section \ref{sec:ofr}. Since the resonance for $\Delta m = 3$ is located at a higher field than the $\Delta m = 5$ case, the final molecule is more deeply bound than the initial one, so the bound-to-bound Raman transition occurs at a higher two-photon detuning than the single particle case (see Fig. \ref{fig:raman_spec}{\bf a}), in particular at $\Delta f = \Delta_{mol} = \Delta_{at}+\Delta E_b$, where $\Delta E_b$ is the binding energy difference between the initial and final molecule and $\Delta_{at}$ is position of the peak corresponding to the $\ket{+5/2}\rightarrow\ket{+1/2}$ transition. \par 
Furthermore, we measured the shift of the bound-to-bound Raman transition peak as a function of the magnetic field at a lattice depth $s=15$. The results of these measurements are reported by the points in the inset of Fig. \ref{fig:raman_spec}{\bf b}. The experimental data are compared to the solid line that represents the theoretical prediction of the binding energy difference between two molecules with $\Delta m = 5$ and $\Delta m = 3$ calculated with our model (see section \ref{sec:photoass}). There is a good agreement between the data and the theory, with some deviations at higher magnetic field values.
\par 
\begin{figure} 
\includegraphics[width=\columnwidth]{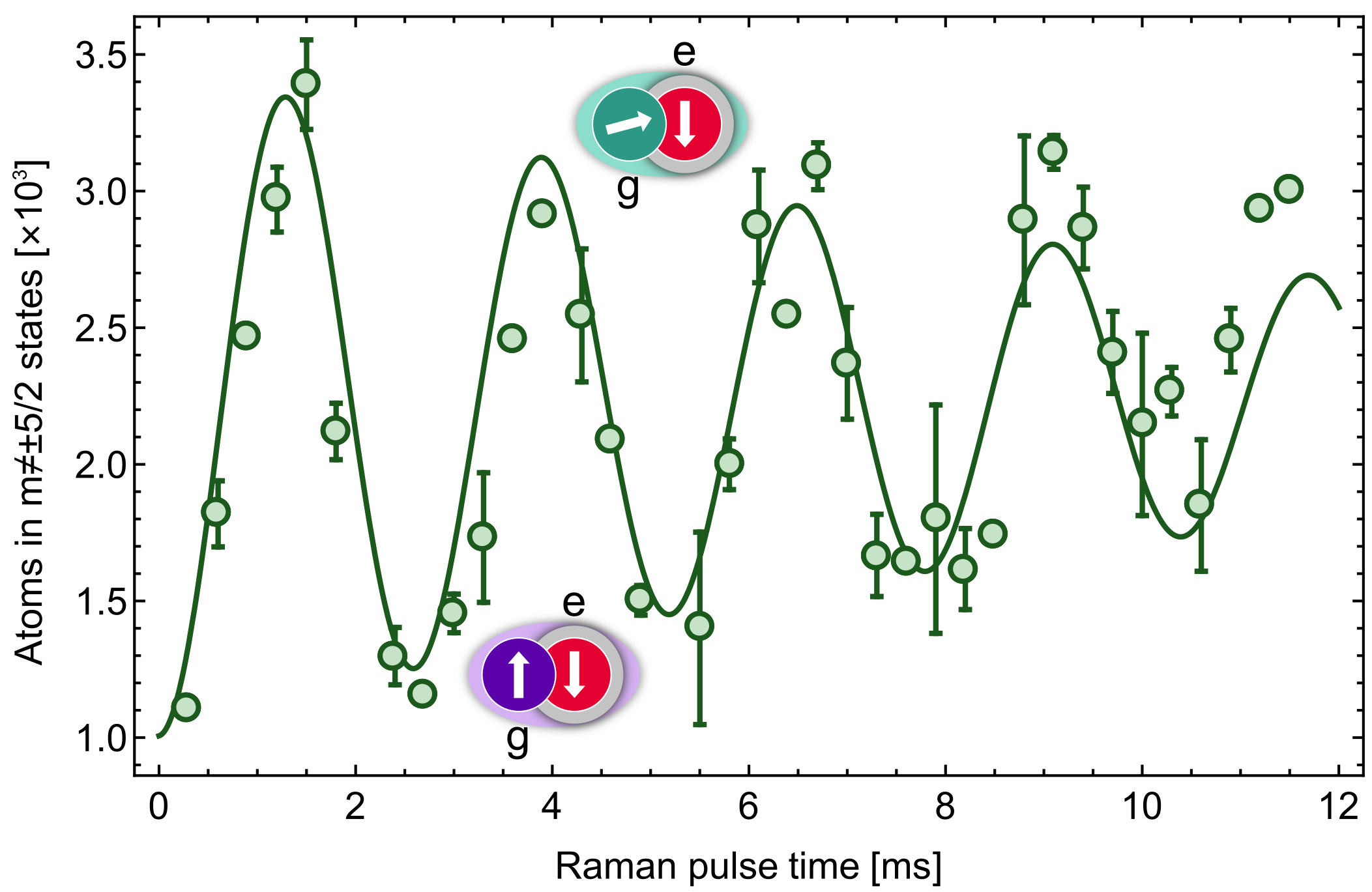}
\caption{Rabi oscillation on the bound-to-bound Raman transition peak performed at $B=150$ G and a 3D lattice depth $s_f=30$. The Raman light coherently flips ths ground-state nuclear spin component of the molecules between the $m=+5/2$ state and $m = +1/2$ state.} \label{fig:raman_osc}
\end{figure}
Similarly to the what we observed for the photoassociation process with the clock laser, this bound-to-bound Raman transition can be driven coherently, as shown in Fig. \ref{fig:raman_osc} where we report a Rabi oscillation performed at a field of $B=150$ G and a 3D lattice depth $s_f=30$. Also in this case it should be noted that this oscillation frequency is fully compatible with that of its single-particle counterpart, showing a Frank-Condon factor which approaches unity.\par

\begin{figure} 
\includegraphics[width=\columnwidth]{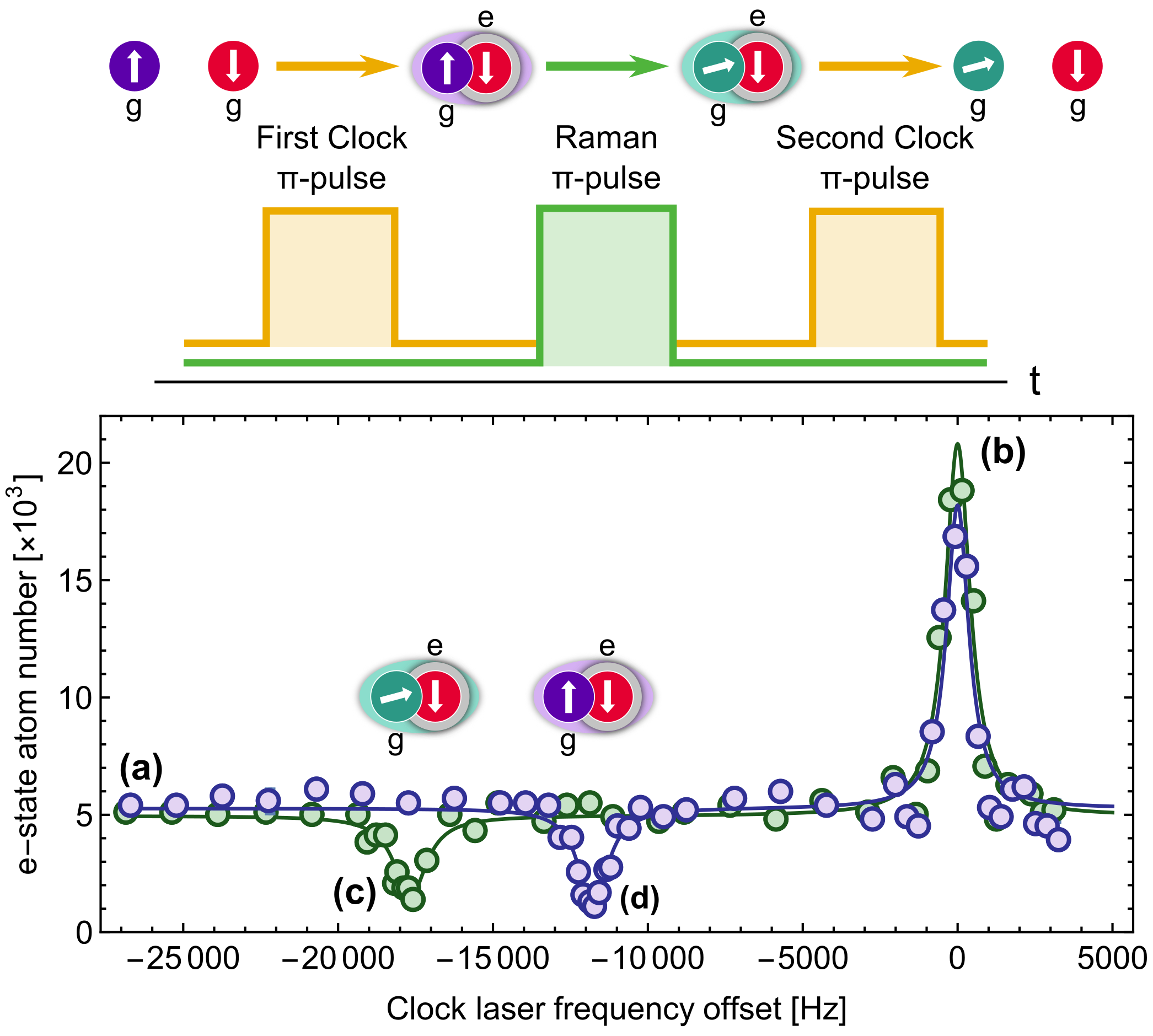}
\caption{Clock laser spectroscopy of a sample with individual atoms and orbital molecules produced with a preliminar clock laser $\pi$-pulse. The excitation of individual atoms is detected as an increase in the number of $\ket{e}$-state atoms ({\bf (b)}) with respect to the background ({\bf (a)}) and serves a zero-frequency reference. The dissociation of orbital molecules is instead detected as a decrease of $\ket{e}$-state atoms ({\bf (c)}). The blue data correspond to an experiment in which the sample is probed immediately after the production of orbital molecules, while the green data correspond to an experiment in which the nuclear spin state of ground-state component of the molecules has been flipped from $+5/2$ to $+1/2$ with a bound-to-bound Raman transition (see text for details). The frequency difference between the two dissociation peaks corresponds to the energy difference between the bound states of the two different mixtures of spin components at 150 G and a $s=30$ lattice depth (see Fig. 2). Both experiments are performed using a sequence of coherent $\pi$-pulses, as can be seen by the nearly full depletion of the photodissociation peaks.} \label{fig:triple_pulse}
\end{figure}

We exploit this property to perform a ``triple-pulse" experiment in which we combine clock-laser photoassociation and Raman manipulation with the goal to verify our capability to manipulate both the orbital ($\ket{g},\ket{e}$) and nuclear-spin degrees of freedom in a coherent way and to further validate the attribution of the bound-to-bound transition peak in Fig. \ref{fig:raman_spec}{\bf b}. Initially, a sample of $m = \pm 5/2$ atoms is prepared in a 3D lattice and orbital molecules are photoassociated in doubly-occupied sites with a clock laser $\pi$-pulse. We then perform a Raman $\pi$-pulse at the bound-to-bound transition frequency so that the $\ket{g,+5/2;e,-5/2}$ molecules are converted into $\ket{g,+1/2;e,-5/2}$ molecules. A second clock laser $\pi$-pulse is then shone on the sample with a frequency that is scanned in order to perform photodissociation spectroscopy. As in the photoassociation spectroscopy experiment of Fig. \ref{fig:photoass_spec} we detect the number of atoms in the $\ket{e}$ state. The results of this experiment for $B=150$ G and $s_f=30$ are represented by the green points of Fig. \ref{fig:triple_pulse}, while the solid green line is a fit to the with a double Lorentzian function. Depending on the frequency of the clock laser during the second $\pi$-pulse, different outcomes are possible. If the second clock pulse is not resonant with any transition (see {\bf (a)} in Fig. \ref{fig:triple_pulse}), the ``background" number of approximately 5$\times 10^3$ $\ket{e}$-state atoms corresponds to the $\ket{e}$-state atoms in the orbital molecules produced with the first $\pi$-pulse. In case instead the second clock pulse is resonant with the excitation of individual $m=-5/2$ atoms in singly-occupied sites, we detect an increase of $\ket{e}$-state atoms (see {\bf (b)}). In analogy with the clock photoassociation spectrum of Fig. \ref{fig:photoass_spec}, the position of this peak is taken as zero-frequency reference. Finally, when the second clock laser pulse is resonant with the molecular peak, the orbital molecules are dissociated to a pair of $\ket{g}$-state atoms and a depletion of $\ket{e}$-state atoms is detected (see {\bf (c)}).
At the center of the resonance, this depletion is noticeably high, with less than $10^3$ atoms that remain in the excited state corresponding to $<20\%$ of the initial number of molecules. For comparison, the blue points and fit in Fig. \ref{fig:triple_pulse} are the result of the same experiment performed without shining the Raman $\pi$-pulse on the bound-to-bound transition, so that orbital molecules are associated with the first clock laser $\pi$-pulse followed immediately by the second clock laser $\pi$-pulse with varying frequency. We detect the same number of background $\ket{e}$-state atoms and a similar dissociation peak, which is located at higher frequency (see {\bf (d)}) with respect to the case in which $\Delta m = 3$ molecules are created by the Raman pulse because of the smaller binding energy of the molecules with $\Delta m=5$. The frequency difference between the dissociation peaks of the two spectra in Fig. \ref{fig:triple_pulse} corresponds to the two-photon detuning difference between the Raman single-particle and bound-to-bound transition peaks of the analogous Raman spectrum of Fig. \ref{fig:raman_spec}{\bf b} performed at $B=150$ G, definitely demonstrating the bound-to-bound nature of the Raman process.

\section{Lifetime of the molecular sample}

In the previous section we showed that the presence of a bound-to-bound transition peak in the Raman spectrum is an unambiguous signature of orbital molecules in the sample. This fact allows for the selective detection of molecules, given that, as already specified before, standard molecules detection techniques cannot be used due to the extremely shallow binding energy of the dimers. In this section, we investigate the lifetime of orbital molecules in different lattice configurations by exploiting this detection method. \par
\begin{figure}
\includegraphics[width=\columnwidth]{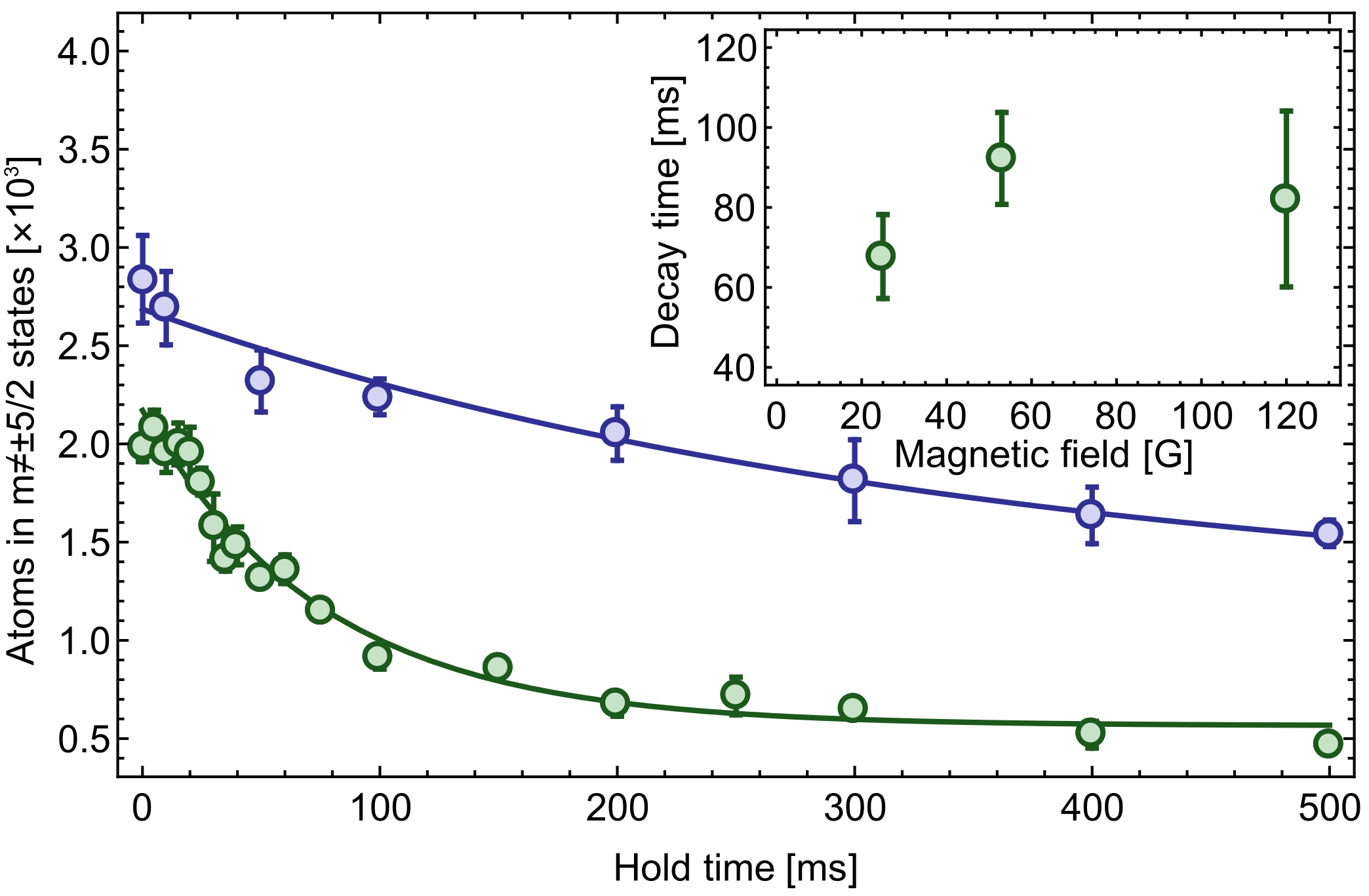}
\caption{Lifetime measurements of orbital molecules in different optical lattice configurations. Blue: lifetime of isolated orbital molecules in a $s_f=15$ 3D lattice. Green: lifetime of interacting molecules in an array of pancakes. In this second scenario, molecules are free to interact with other atoms and molecules of the sample, leading to a shorter lifetime with respect to the 3D lattice case. The inset shows the measured lifetime of orbital molecules in pancakes for different values of magnetic field across the orbital Feshbach resonance. The lifetimes values are compatible within the error bars and no evident trend is observable.} \label{fig:decays}
\end{figure}

In a first experiment, we study the lifetime of isolated molecules confined in a 3D optical lattice. This is done by photoassociating molecules with a clock laser $\pi$-pulse in a 3D lattice of depth $s_f$ and at a magnetic field $B$ as in the previous experiments. The sample is then held in the 3D lattice for a variable hold time, and a Raman $\pi$-pulse resonant with the bound-to-bound two-photon transition is performed after the hold time in order to detect the number of molecules. The blue data in Fig. \ref{fig:decays} correspond to the result of this measurement as a function of the hold time in a $s_f=15$ 3D lattice performed at $B=25$ G. We fit the data with an exponential curve obtaining a lifetime of the isolated molecules of $350(70)$ ms, a remarkably long value despite one of the atoms forming the molecule being in a highly-excited state. The main limiting factor to the lifetime could be identified as inelastic losses due to collisions after a tunneling process to an occupied neighboring site (the tunneling rate is of the order of 10 Hz). \par

In a similar experiment, we use this technique to carry out a first study of orbital molecules lifetime in a many-body environment. After the clock photoassociation of molecules in the 3D lattice, the horizontal lattice beams are turned off with a 1-ms-long linear ramp to obtain a vertical stack of pancakes with a radial trapping frequency of approximately $2\pi\times 20$ Hz and a strong axial confinement of $2\pi\times 15$ kHz, corresponding to an optical lattice depth of $s=15$ along the vertical direction. The sample is held in this pancakes configuration for a variable hold time, then the horizontal lattices are turned on again with 1-ms-long linear ramps and a Raman $\pi$-pulse is performed to detect the number of remaining molecules as in the previous experiment. The results of such an experiment performed at a field of 25 G is reported by the green data in Fig. \ref{fig:decays}, where it is compared to the lifetime of isolated orbital molecules in the 3D lattice. In the pancakes, the molecules are free to interact both with other molecules and single ground-state atoms in the $m=\pm 5/2$ states. The decay rate of the molecules could than be described by a rate equation $\dot{n}_m=-\beta_{ma}n_mn_a-\beta_{mm}n_m^2$, where $n_m$ and $n_a$ are the density of molecules and individual atoms, respectively, and $\beta_{ma}$ and $\beta_{mm}$ are the loss rate coefficients of the molecule-atom and molecule-molecule loss processes, respectively. In our experimental conditions, the typical number of molecules (a few thousands) is much smaller than the typical number of atoms (typically of the order of $50\times 10^3$), so it is reasonable to assume that the dominant loss mechanism is represented by molecule-atom collisions. In this case, the second term of the right hand of the previous rate equation can be neglected, and the model reduces to a simpler rate equation $\dot{n}_m=-\beta_{ma}n_mn_a$,
which is solved by a single exponential curve $n_m = e^{-\beta_{ma}n_at} = e^{-\gamma_{ma}t}$. We fit the green data of Fig. \ref{fig:decays} with an exponential function obtaining a lifetime of $1/\gamma_{ma}=77(8)$ ms. In order to quantify the loss rate coefficient related to this process, we use a simplified model to calculate the average density in every pancake starting from our initial sample with $N$ atoms and $T/T_F = 0.25$, similarly to the model used to calculate the density of bosonic $^{174}$Yb in \cite{Franchi2017}. We then average over all the pancakes to obtain a mean density value of $n_{at}=6.3(3.1)\times 10^{12}$ cm$^{-3}$, to which we attribute a conservative error due to the several assumptions in our theoretical model. From this value we can determine a loss rate coefficient $\beta_{ma} = \gamma_{ma}n_{a} = 2.1(1.2)\times 10^{-12}$ cm$^{3}$s$^{-1}$. \par
This method allowed us to measure the molecules lifetime in the sample for different values of magnetic field all across the free-space orbital Feshbach resonance. The results of these measurements are reported in the inset of Fig. \ref{fig:decays}. Each point is the average of two independent experiments performed in different days and the error bar is the average of the fit errors of the individual experiments. Noticeably, there is not a clearly visible trend and the measured lifetimes are compatible within the error bars, independently on the magnetic field being set below or above or near resonance. This is in general not true for Feshbach molecules of fermionic alkali atoms, which show faster losses on the BEC side of the resonance and a longer lifetime near resonance arising from the Pauli exclusion which suppresses collisions between the fermionic constituents of the weakly-bound dimers. This fact is a further suggestion that our lifetime measurements are dominated by molecules-atom losses, which are not protected by the Pauli exclusion principle due to the presence of atoms in a ``third" state (the $\ket{g,-5/2}$ atoms in the singly-occupied sites of the initial 3D lattice) different from both the states of atoms forming the $\ket{g,+5/2;e,-5/2}$ orbital molecule. A more detailed investigation of the multiple interaction mechanisms, especially of molecule-molecule scattering, would require the creation of pure samples of orbital molecules, a very difficult task in Yb samples due to the small binding energy of the molecules which makes it impossible to selectively address and blast from the sample single atoms exploiting ``non-clock" transitions. Nevertheless, the measured lifetimes of the order of 100 ms are a promising starting point for future studies of the BEC-BCS crossover in this unexplored regime, as it represents a lower limit to the lifetime that could be obtained in pure molecular samples. \par

\section{Conclusions}

In this paper we have demonstrated the possibility to produce and coherently manipulate Feshbach molecules made by two-electron atoms in different electronic ``clock" states $^1$S$_0$ and $^3$P$_0$, and provided a first investigation of the molecule lifetime in a many-body setting. We have shown how these molecules can be produced in a three-dimensional optical lattice by means of direct optical photoassociation with clock laser light and have demonstrated that this photoassociation process can be driven coherently, leading to several cycles of association and dissociation. Noticeably, the achievable Rabi frequencies are comparable to that of the excitation of a single-atom clock transition under the same experimental conditions, highlighting a favorable Frank-Condon factor. This opens new possibilities towards the realization of molecular optical clocks \cite{Borkowski2018}, as well as for the exploration of new approaches for the control of interactions in two-electron atoms with optical Feshbach resonances \cite{optical1,optical2,optical3}.
\par
We also demonstrated that the molecular internal degree of freedom can be controlled exploiting Raman transitions that swap between different nuclear spin states of the $^1S_0$ atom forming the molecule. We show that also this process can be driven coherently, allowing us to cycle between molecular states with different ground-state component and providing a very powerful manipulation tool, which could also be exploited for the optical tuning of an optical Feshbach resonance \cite{optical4}. Moreover, we showed that this manipulation scheme can also be used as a detection tool for orbital molecules, which are characterized by an extremely shallow binding energy that causes the usual detection methods of Feshbach molecules of alkaline atoms to be unusable. This allowed us to perform a first investigation of orbital molecules in a many-body environment, where we measured lifetimes approaching 100 ms. This value, that is already of the same order of the typical experimental timescales, is particularly promising since it appears to be mainly limited by the presence of a large number of unpaired atoms in the sample, and is therefore a lower limit to the lifetime that could be achieved in a pure sample of molecules. In future works, the possible development of a ``purification" method and improved molecule production techniques could allow for the creation of pure and larger samples of molecules leading to further and more detailed experiments to characterize the different scattering processes of orbital molecules.  potentially opening the way to the investigation of the BEC to BCS crossover in atoms with two internal degrees of freedom.

\section*{Acknowledgements}

We are very grateful to H. Zhai, P. Zhang, R. Zhang and Y. Cheng for the numerous inspiring discussions and for suggesting us to perform Raman measurements on orbital molecules. We also acknowledge insightful discussions within the LENS QuantumGases group, in particular with G. Roati, F. Scazza and M. Zaccanti. We thank D. Calonico (INRIM) for borrowing us the clock laser chip, and TOPTICA Photonics AG for their prompt and high-quality technical assistance. \par
We acknowledge financial support from ERC Consolidator Grant TOPSIM, QuantERA project QTFLAG, INFN project FISH, MIUR project FARE TOPSPACE and MIUR PRIN project 2015C5SEJJ.

\end{document}